\documentclass[journal,10pt,twocolumn,twoside]{IEEEtran}	 
\abovedisplayshortskip \abovedisplayskip \relax
\ifCLASSINFOpdf
\else
\fi

\usepackage{graphicx}
\usepackage{subfigure}
\usepackage{epstopdf}
\usepackage{amsfonts}
\usepackage{amsmath}
\usepackage{amssymb}
\usepackage{wasysym}
\newcommand{\argmin}{\operatornamewithlimits{argmin}}

\begin{document}
\title{Combination of Linear Prediction\\and Phase Decomposition\\for Glottal Source Analysis on Voiced Speech}

\author{Yiqiao~Chen and John~N.~Gowdy \thanks{The authors are with Department of Electrical and Computer Engineering, Clemson University
(e-mail: yiqiaoc@clemson.edu; gowdy@clesmon.edu).} }

\markboth{IEEE SIGNAL PROCESSING LETTERS}%
{CHEN \MakeLowercase{\textit{et al.}}}
\maketitle
\begin{abstract}
Some glottal analysis approaches based upon linear prediction or complex cepstrum approaches have been proved to be effective to estimate glottal source from real speech utterances. We propose a new approach employing both an all-pole odd-order linear prediction to provide a coarse estimation and phase decomposition based causality/anti-causality separation to generate further refinements. The obtained measures show that this method improved performance in terms of reducing source-filter separation in estimation of glottal flow pulses (GFP). No glottal model fitting is required by this method, thus it has wide and flexible adaptation to retain fidelity of speakers's vocal features with computationally affordable resource. The method is evaluated on real speech utterances to validate it.

\end{abstract}

\IEEEpeerreviewmaketitle

\section{Introduction}

\IEEEPARstart{V}{oiced speech} is typically modeled as the output of source-filter model with glottal flow pulses as excitation source. The estimation of glottal flow pulses (GFPs) is equivalent to removing vocal-tract filter (VTF) components from the observed speech utterance. VTF responses are normally considered to be contributed by formants or resonances from vibration of vocal cords in terms of speakers. Consequently, the separation could be achieved by estimating the VTF response and its further cancellation from the speech. Linear prediction (LP) and inverse filtering have being played important roles in some source estimation algorithms [1], [2]. These methods find model parameters for a specified discrete all-pole (DAP) modeling and a predetermined number of parameters. This method lacks the accuracy to restore the GFPs. 

Some other experiments also demonstrated that glottal source or its open phase [3], [4] is contributed by maximum-phase components, and VTF response is contributed by poles and lossy zeros inside unit circle [5]. Based that information, phase decomposition can also be made to achieve the source-filter separation. Some earlier polynomial factorization methods [7] including zeros of $\mathcal{Z}$-transform (ZZT) and complex cepstrum (CC) approaches [8] work towards the decomposition based on the windowed finite-length speech sequence within two consecutive glottal closure instants (GCIs). They are efficient and free of any potentially biased usage of hypothesized glottal models, though effects of vocal-tract resonances could remain after the decomposition.

To combine the advantages of the two widely used categories of estimation strategies, we thus presented a jointly linear-prediction and complex cepstrum (LPCC) approach employing a non-classical odd-order DAP model and CC phase decomposition. The presented approach has good qualitative and quantitative performance through validations of both waveform shape and parameterization of Liljencrant-Fant (LF) models in terms of voiced speech utterances. It provides a straightforward way to discard VTF effects from speech and recover a clean glottal source for each voiced pitch-long speech without any hypothesized glottal source model.

\section{Problem Formulation}
The glottal wave estimation can be considered an inverse problem, in which the derivative exciting signal $G(z)$ needs to be derived from an observed voiced speech signal $X(z)$ for  source-filter model in the classical speech production mechanism
\begin{equation} 
X(z) = G(z)H(z)L(z)
\end{equation}
where $H(z)$ and $L(z)$ denote the VTF and lips radiation. The problem is normally dealt with by source-filter separation. Because real glottal flows as the excitation to VTF are always unknown, phase decomposition methods avoid specific assumption about excitation model shape within each shifting analysis window.
 
Let $w[n]$ be the truncated signal within a window and $w[n]\overset{\mathcal{Z}}{\longleftrightarrow}W(z)$ where
\begin{equation*}
W(z)=1+\sum\limits_{i=1}^{N} w_i z^{-i}=\prod\limits_{i=1}^{N} (1-q_i z^{-i}),~1\leq{i}\leq{N}.
\end{equation*}
It is difficult to suppress infinite impulse responses (IIR) from complex conjugate pole pairs, especially those close to unit circle with slow attenuation in time domain, merely by removing finite minimum-phase zeros from the polynomial $W(z)$. Thus, LP inverse filtering [1] with complex poles has an irreplaceable advantage to suppress these resonances from VTF. Meanwhile, at least one real pole is required to fit the spectral representation of an individual glottal pulse [9]. That can be guaranteed from an odd-order LP analysis rather than commonly used even-order LP analysis. We then propose the method in next section being motivated by these points to enhance GFP estimate performance by combining the LP analysis and phase-decomposition with complex cepstrum.

\section{Proposed Approach}
%
%

\subsection{Short-Time Vocal-Tract Estimation and Inverse Filtering}
All-pole LP analysis has been a major tool adapted by some earlier methods [1], [2] to glottal source analysis as a short-time estimation to the speaker's varying vocal-tract response. With an odd-order predictor, there always exists at least one real root and pairs of complex conjugate roots corresponding to source and VTF components respectively. A coarse estimation of glottal source can be obtained by inverse filtering with only complex pole pairs. 
   
Applying a short-time hamming window with lenght of $N$ to the input signal $X(z)$, we obtain a coefficient vector
$\bf{a}$$^{(n)}  \triangleq {[~a^{(n)}_1~a^{(n)}_2~\cdots~a^{(n)}_{2M+1}~]^{^T} \in \mathbb{R}^{2M+1}}$ which can be found by covariance method for the current all-pole model
\begin{equation}
\hat{H}_{ap}^{(n)}(z) = \frac{1}{1-\sum\limits_{m=1}^{2M+1} a^{(n)}_m z^{-m}}
\end{equation}
\noindent with recursive least-squares (RLS) algorithms to solve

$$\mathop{\argmin}_{{\bf{a}}^{(n)}}{\|{\bf{x}}^{(n)}-\tilde{\bf{X}}^{(n)}{\bf{a}}^{(n)}\|}$$

\noindent where data matrix $\tilde{\bf{X}}^{(n)} \in \mathbb{R}^{N\times{(2M+1)}}$ is formed from shifted versions of the current observation data frame ${\bf{x}}^{(n)}$. Excluding all real poles from $\hat{H}_{ap}^{(n)}(z)$, we have
\begin{equation}
\hat{H}^{(n)}(z)=\frac{1}{\prod\limits_{i=1}^{\ell} (1-p^{(n)}_i z^{-i})(1-{p^*_i}^{(n)} z^{-i})}\end{equation}
as the short-time estimation of $H^{(n)}(z)$ on the current frame with $2\ell<M$. 

The spectral representation of a VTF with 10th-order LP and another one with 11th-order LP with respect to a pressed vowel /u/ are shown in Fig. 1 where a real pole is generated by the odd-order LPC analysis in an 8kHz sampling rate environment.

An estimation of glottal source signal $\hat{G}^{(n)}(z)$ on the current frame is then obtained by inverse filtering $\hat{X}^{(n)}(z)/\hat{H}^{(n)}(z)$ as well as lips' cancellation. Fig. 2 shows a comparison of $\hat{G}^{(n)}(z)$ with a corresponding accompanied real EGG signal. The coarse estimation here provides a foundation for further processing in next step.

%
%
%
%

\begin{figure}
\centering  
\includegraphics[scale=0.78]{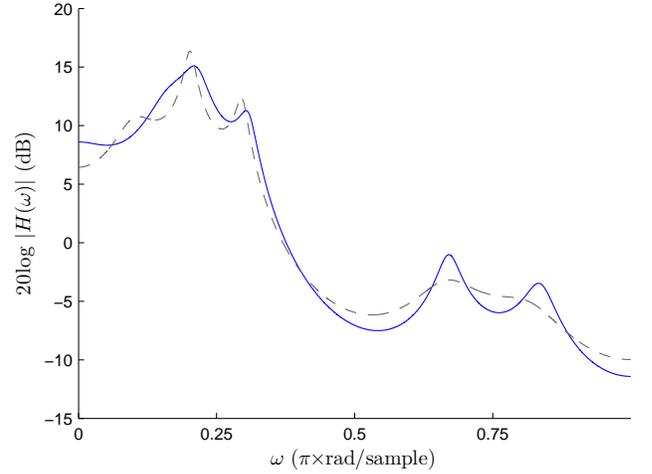} 
\caption{Spectrum comparison of VTF shapes predicted by 10th-order all-pole model in gray dash and 11th-order all-pole model in blue solid}             
\end{figure}
\begin{figure}
\centering  
\includegraphics[scale=0.84]{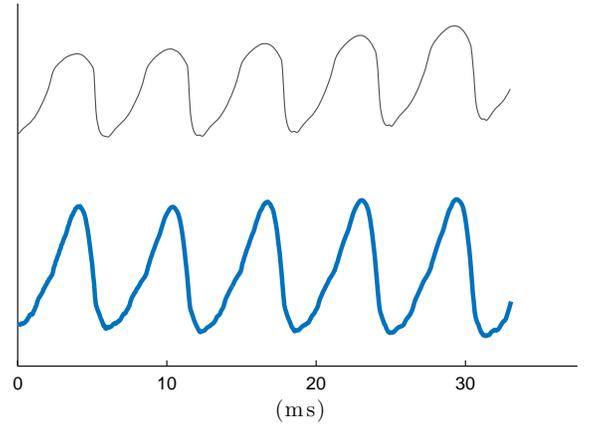} 
\caption{Estimated GFP after inverse filtering on a recorded utterance from a real male speaker and its EGG are shown in thick blue and thin gray.}             
\end{figure}
%


\subsection{Complex Cepstrum Processing}
Removing residual sinusoidal components after LP inverse filtering by the DAP model $\hat{H}^{(n)}(z)$ in (3), we are able to get refined derivative pulses based upon those calculated GCI locations by similar differentiation methods [11] on EGG signals. Then homomorphic filtering can be employed to separate anti-causality from causality signal upon the coarse estimates.

Let $e[n]$ be a truncated coarse glottal source estimation $\hat{G}^{(n)}(z)$ slightly larger than pitch-length by choosing a suitable analysis region between neighborhoods of two consecutive GCIs $s_{k}+\epsilon_{k}$ and $s_{k+1}+\epsilon_{k+1}$ to achieve the best performance for current pulse [18]. Homomorphic filtering can then be applied to $e[n]$ described as combination of minimum and maximum-phase components 
\begin{equation}
e[n]=e_{\mathrm{min}}[n]\ast{e_{\mathrm{max}}[n]}
\end{equation}

After phase unwrapping and determination of the algebraic sign of the gain $A$ of $E(z)$, the computation of the finite-length CC of 
${\tilde{e}}[n]$ leads to the cepstrum [10]\\
\begin{equation}
\hat{c}_{e}[n]=\mathcal{Z}^{-1}\{\ln{E(z})\}=\left\{ \begin{aligned}
\ln|A|,  \quad n=0\\
-\sum\limits_{m} \frac{a_m^n}{n},  \quad n>0\\
\sum\limits_{k} \frac{b_k^n}{n},  \quad n<0
\end{aligned}\right.
\end{equation}
\\
\noindent where coefficients $a_m$ and $b_k$ are listed as polynomials minimum-phase roots and maximum-phase roots’ reciprocals respectively. 

\begin{figure}
\raggedleft 
\includegraphics[scale=0.62]{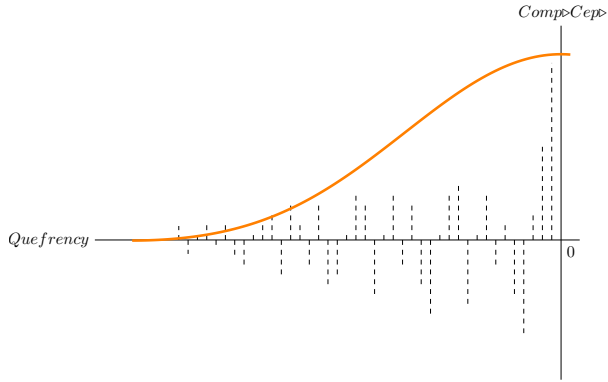} 
\caption{Anti-causal part with half-long window attenuation}             
\end{figure}

The minimum-phase portion was dropped as other phase decomposition methods do. Additionally, some distortions leaked into maximum-phase portion of cepstrum need to be taken care of due to CC domain's aliasing from periodic extension in terms of DFT computation after logarithmic operation. A half-length window function with an attenuation in high quefrency components of anti-causal region is thus utilized to suppress the aliasing effects as Fig. 3 illustrates before transforming back to time domain.



\section{Experimental Results}



\begin{figure*}[htb]

\centering                           
\includegraphics[scale=0.7]{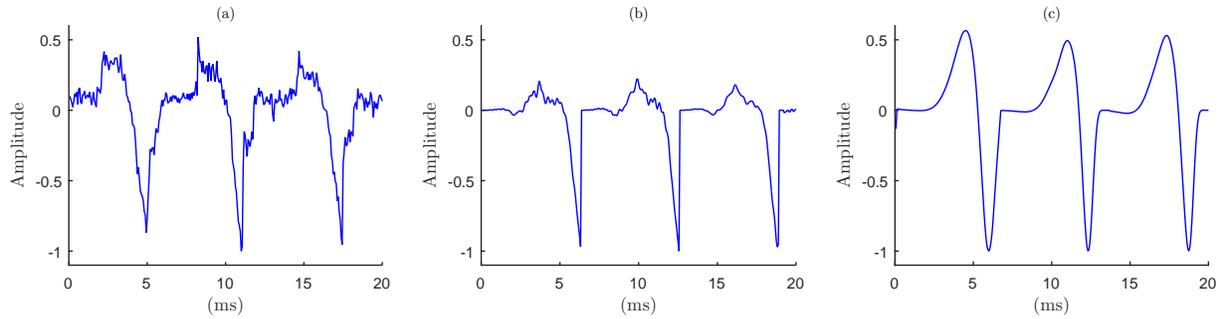}
\caption{Results of estimation for a fragment of /{\inve}/ : (a) Estimated glottal waveform with IAIF. (b) Estimated glottal waveform with CC. (c) Estimated glottal waveform with LPCC.}
\end{figure*}

\subsection{Application of Real Speech Waveforms} 
Real speech utterances of vowel sounds from three English speakers, two males and one female, were analyzed by the proposed approach and other existing ones. The LPCC generates estimated source waveform with generally less distortions than other CC and DAP-based approaches. Fig. 4 demonstrates an experiment where a 25ms short-time shifting window was applied to a recorded real male sound /i/ sampled in 16kHz. Here a 19th-order DAP model was employed for LP analysis and inverse filtering with estimated resonance response and lip's radiation first, then a half-length blackman window was applied to the maximum-phase quantities on the left of origin of cepstrum after discarding minimum-phase components. Instances of the estimated pulses along with results by CC and IAIF are also shown in the experiment. One can see that the ripples in the estimated glottal source are remarkably significant with the IAIF approach in Fig. 4(a) because of the lack of precision of the DAP method. Noticeable ripples and damped effects in CC methods are also obvious in Fig. 4(b) because of distortion from residual VTF resonance. Compared to other approaches, the proposed approach in Fig. 4(c) eventually offers a clean derivative waveform with pitch-by-pitch analysis after executing the two steps described, from which resonances from vocal-tract were removed. 
  
\subsection{Comparison with Electro-Glotto-Graphic Signals}
EGG is a tool for the noninvasive measurement of the time variation of the degree of contact between the vibrating vocal folds during voice generation. Timing features of given speech utterances can be determined through EGG signals. These features include the calibrated locations of instants of closure, opening of the glottis and the calculated open quotient $O_q$ in voiced segments of EGG signals. As a reference, the calculated open quotients $O_q$ can be regarded as the ground truth to compare to features measured and calculated from estimated glottal source by different analysis approaches.

Variations of the speech signal will lead to the variations of its corresponding dEGG signal and the glottal source derivative estimated by LPCC. It forms the foundation of further comparisons. The dEGG and its derived GCIs through the DECOM method [11] in Fig. 5(b) for a human spoken sequence is shown in Fig. 5(a). Fig. 5(c) shows the estimated derivative glottal source by LPCC after shape normalization. The sequence of pulses were concatenated and arranged by phase decomposition window and LP window locations. The valleys in this graph symbolized as GCIs must be aligned to obtained derivative signal after applying DECOM to GFPs in Fig. 2. Given the closure instants, we need to indicate the opening counterpart for each pitch period. We adapted LF parameters like former papers [6], [14], [15], [16] to derive $O_q$ of glottal estimate. Fitting to any individual estimated source derivative, LF's $t_o$ and $t_e$ serving as opening and closure instants can be used to determine the opening-phase $(t_e - t_o)$, the nominator to derive $O_q$ together with measured pitch length of the current pulse. Obviously, the accuracy of $O_q$ depends on accurate estimates of both $t_e$ and $t_o$.

Given EGG signal from the 16 kHz $bdl$ and $jmk$ CMU $Arctic$ database, values $t_e$ and $t_o$ instants can be measured in by DECOM, then $O_q = f_0(t_e - t_o)$. We then evaluate the proposed approach through comparing $O_q$ values. The estimated open-quotient values by all approaches have at least 0.1 as an estimation error from the measured values in EGG signals as an example shown in Fig. 6.

Several factors result in errors' fluctuation of open quotients estimation. The choice of LF model itself might also lead to errors in some extent as it is just an approximation of real glottal flows. More important, a larger fluctuation of ripples in estimated glottal source brings bigger fitting errors and corresponding troubles to gradient searching in LF fitting which relies on the optimization of nonlinear cost functions. Consequently, rising variances in opening instants in LF calculation will add instabilities of opening instants and $O_q$ even though determinations GCIs are less flexible because they were determined before the fitting. These approaches have similar error mean values. And LPCC outperforms other two methods by $O_q$ estimation error's smallest variance and stability in the fitting process. It is due to a higher quality of estimation in terms of glottal derivative's shape estimation in order to eliminate vocal-tract's resonance effects described in the last experiment.

\begin{figure}
\centering     
\includegraphics[scale=0.9]{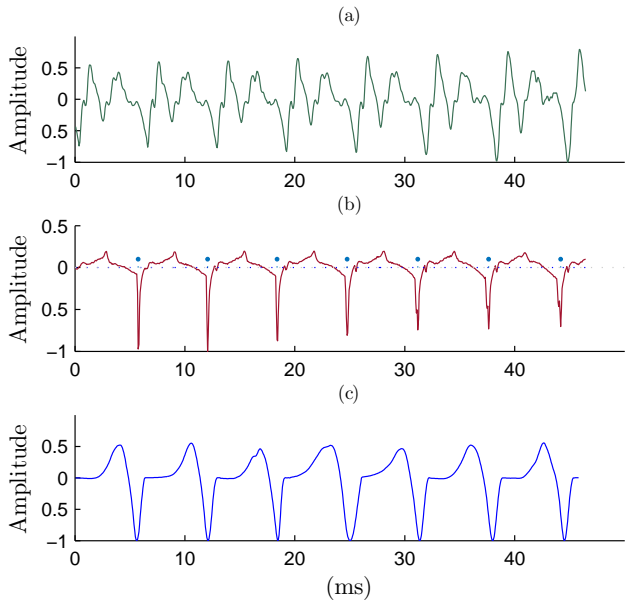} 
\caption{ GCI/Glottal derivative estimations on recorded from \textit{bdl} CMU \textit{Arctic} database: (a) Utterance fragment of  ''Author of ...''. (b) dEGG signal of the utterance in (a). (c) Estimated derivative pulses with 19-pole LPCC approach.}             
\end{figure}
\begin{figure}
\raggedleft		
\includegraphics[scale=.8]{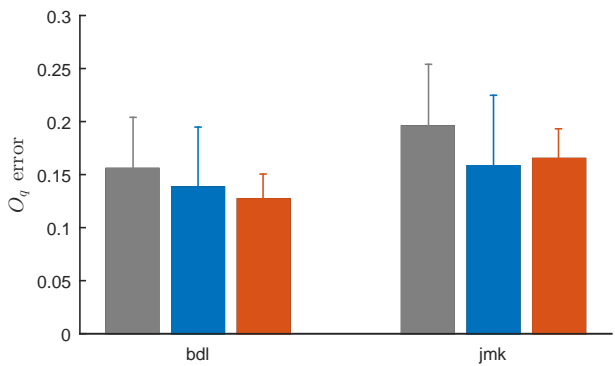} 
\caption{ Estimation $O_q$ errors of some approaches. From left to right: IAIF, CC, LPCC.}
\end{figure}

\section{Conclusion}
Utilizing revised DAP and phase decomposition approaches, we have presented an LPCC approach to successfully recover glottal flows for voiced sounds in this letter. It has a big advantage to generate clean waveform estimates that can be potentially explored for future applications [14] requiring speaker's accurate glottal features. In terms of open-quotient estimation, LPCC is more reliable than IAIF and CC due to better waveform fitting to parameter-adjustable LF model because the less ripples in waveform lead to less error in the fitting process.  

\ifCLASSOPTIONcaptionsoff
  \newpage
\fi


\begin{thebibliography}{99}

\bibitem{IEEEhowto:Alku}
P.~Alku, “Glottal wave analysis with pitch synchronous iterative adaptive inverse filtering,”
	\textit{Speech Commun}., vol. 11, no. 2--3, pp. 109--118, 1992.

\bibitem{IEEEhowto:Wong}
D.~Wong, J.~D. Markel, and A.~H. Gray, “Least squares glottal inverse filtering from the acoustic speech waveform,”
	\textit{IEEE Trans. Acoust., Speech, Signal Process}., vol. ASSP--27, no. 4, pp. 350--355, Aug. 1979.

\bibitem{IEEEhowto:Rosenberg}
A.~Rosenberg, “Effect of the glottal pulse shape on the quality of natural vowels,”
	\textit{J. Acoust. Soc. Amer},  vol. 49, pp. 583--590, 1971.

\bibitem{IEEEhowto:Fant}
G.~Fant, “The LF-model revisited. transformations and frequency domain analysis,” \textit{STL-QPSR}, vol. 36, no. 2-3, pp. 119–-156, 1995.

\bibitem{IEEEhowto:Doval}
G.~Doval, C.~D'Alessandro, and N.~Henrich, “The voice source as a causal/anticausal linear filter,”	\textit{VOQUAL}, 2003.

\bibitem{IEEEhowto:Degottex}	
G. Degottex, A. Roebel and X. Rodet, “Phase minimization for glottal model estimation,”
	\textit{IEEE Trans. Audio, Speech, Lang. Process}., vol. 19, no. 5, pp. 1080--1090, Jul. 2011.

\bibitem{IEEEhowto:Drugman}
T. Drugman, B. Bozkurt and T. Dutoit, “Causal-anticausal decomposition of speech using complex cepstrum for glottal source estimation,”
	\textit{Speech Comm}., vol. 53, no. 6, pp. 740--741, Jul. 2011.

\bibitem{IEEEhowto:Bozkurt}
B. Bozkurt, B. Doval, C. D'Alessandro, and T. Dutoit, “Zeros of z-transform representation with application to source-filter separation in speech,” 		
	\textit{IEEE Signal Process. Lett}., vol. 12, no. 4, pp. 344--347, Apr. 2005.

\bibitem{IEEEhowto:Quateri}
T.~Quatieri, Discrete-Time Speech Signal Processing: Principles and Practice. Upper Saddle River, NJ: Prentice Hall, 2001.

\bibitem{IEEEhowto:Oppenheim}
A.~Oppenheim and R.~Schafer, Discrete-Time Signal Processing, 3rd ed., Upper Saddle River, NJ, Prentice Hall. 2009.
	
\bibitem{IEEEhowto:Henrich}	
N. Henrich, C. D'Alessandro, B. Doval, and M. Castellengo, “On the use of the derivative of electroglottographic signals for characterization of nonpathological phonation,” 
    \textit{J. Acoust. Soc. Amer.}, vol. 115, no. 3, pp. 1321–-1332, 2004.	
	
\bibitem{IEEEhowto:Smits}	
A. Kounoudes, P. A. Naylor, and M. Brookes, “The DYPSA algorithm for estimation of glottal closure instants in voiced speech,”
	\textit{Proc. ICASSP}, pp. I-349–I-352, 2002.
	
\bibitem{IEEEhowto:Thomas}
M.~Thomas and P.~Naylor, “The sigma algorithm: A glottal activity detector for electroglottographic signals,”
	\textit{IEEE Trans. Audio, Speech, Lang. Process}., vol. 17, no. 8, pp. 1557--1566, Nov. 2009.

\bibitem{IEEEhowto:Plumpe}	
M. D. Plumpe, T. F. Quatieri, and D. A. Reynolds, “Modeling of the
glottal flow derivative waveform with application to speaker identification,”
	\textit{IEEE Trans. Speech Audio Process}., vol. 7, no. 5, pp. 569--576, Sep. 1999.
	
\bibitem{IEEEhowto:Degottex}
G. Degottex, J. Kane, T. Drugman, T. Raitio and S. Scherer, “COVAREP - A collaborative voice analysis repository for speech technologies,” 
    \textit{Proc. ICASSP}, 2014.

\bibitem{IEEEhowto:Cab}    	
J. P. Cabral, K. Richmond, J. Yamagishi and S. Renals, “Glottal Spectral Separation for Speech Synthesis,”
	\textit{IEEE Journal Sig. Proc.}., vol. 8, no. 2, pp. 195--208, Apr. 2014.	

\end{thebibliography}
\end{document}